\pdfoutput=1
\documentclass{ifacconf}

\usepackage{graphicx}    
\usepackage{natbib}        
\usepackage{url}
\usepackage{textcomp}
\usepackage{gensymb}
\usepackage{enumitem,kantlipsum}
\usepackage{mathtools}
\usepackage{tikz}
\usetikzlibrary{arrows,positioning} 
\usetikzlibrary{chains, positioning} 
\usetikzlibrary{shapes.geometric, arrows, positioning, calc, matrix}
\usetikzlibrary{shapes.geometric,backgrounds}

\usepackage{tcolorbox}

\usepackage{algorithm}
\usepackage{algpseudocode}
\usepackage{graphicx}
\usepackage[export]{adjustbox}
\usepackage{caption}
\usepackage{subcaption}
\usepackage{graphicx, caption}
\usepackage{lipsum} 
\usepackage{multicol}

\begin{document}

\begin{frontmatter}
\title{Stereo Vision for Unmanned Aerial Vehicle Detection, Tracking, and Motion Control} 
\author[First]{Maria N. Brunet} 
\author[Second]{Guilherme Aramizo Ribeiro} 
\author[Third]{Nina Mahmoudian}
\author[Fourth]{Mo Rastgaar}
\address[First]{School of Electrical and Computer Engineering, Purdue University, West Lafayette, IN 47906 USA (e-mail: mbruneta@purdue.edu)}
\address[Second]{Polytechnic Institute, Purdue University, West Lafayette, IN 47906 USA (e-mail: garamizo@purdue.edu)}
\address[Third]{School of Mechanical Engineering, Purdue University, West Lafayette, IN 47906 USA (e-mail: ninam@purdue.edu)}
\address[Fourth]{Polytechnic Institute, Purdue University, West Lafayette, IN 47906 USA (e-mail: rastgaar@purdue.edu)}
\begin{abstract}                
An innovative method of detecting Unmanned Aerial Vehicles (UAVs) is presented. The goal of this study is to develop a robust setup for an autonomous multi-rotor hunter UAV, capable of visually detecting and tracking the intruder UAVs for real-time motion planning. The system consists of two parts: object detection using a stereo camera to generate 3D point cloud data and video tracking applying a Kalman filter for UAV motion modeling. After detection, the hunter can aim and shoot a tethered net at the intruder to neutralize it. The computer vision, motion tracking, and planning algorithms can be implemented on a portable computer installed on the hunter UAV.

\end{abstract}
\end{frontmatter}

\section{Introduction}
Unmanned Aerial Vehicles (UAVs), in particular multi-rotor UAVs, keep increasing in number. Due to reduced prices, increased availability, improvements in technology, and creation of developer communities, numerous applications have been opened up, but that also led to the misuse of this technology. Concerns keep arising as commercial UAVs can be easily misused by carrying harmful material or performing unwarranted reconnaissance. As a consequence, there is a need for systems that can safely mitigate the threat that UAV misuses may pose to the public safety. For these cases, using a UAV hunter to intercept, capture, and remove intruder UAVs to a safe zone for neutralization is proposed. 

A hunter prototype that can capture an intruder UAV in a tethered net and drag it to a designated location was developed by \cite{aagaah_drone_2018}. 
 
This paper presents the development and evaluation of a visual-based UAV tracking algorithm and a UAV motion controller for the autonomous aiming of the hunter UAV.

Related work in UAV detection shows the use of acoustic signals, radar 
or video analysis methods \cite{schumann_deep_2017}. As for video tracking for aerial vehicles, an approach using optical flow can be found in \cite{krukowski_tracking_nodate}. 
For the motion control of UAVs, \cite{lee_geometric_2010} presents the tracking control of a quadrotor.

The remainder of the paper is organized as follows. In Section 2, the vision system is described and the detection, tracking and motion planning algorithms are presented. Section 3 extends the evaluation of these algorithms in a computer-simulated environment. Section 4 comprises the review of the proposed methods and future work.

\section{UAV Targeting Algorithm}
The UAV Targeting Algorithm involves: 1) Reconstructing the surrounding world using computer vision, 2) Detecting the intruder UAV at each time instant, 3) Tracking the intruder UAV from a time history of the detections, and finally 4) Controlling the position of the hunter UAV to aim the net launcher at the intruder.

\begin{figure}[b]
    \begin{center}
        \includegraphics[width=8cm]{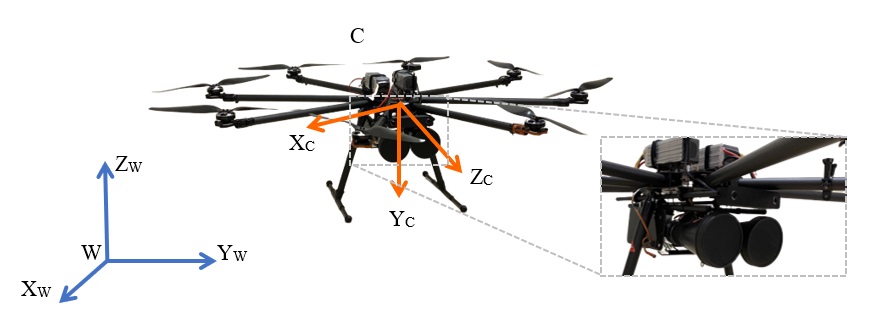}
        \caption{Coordinate frames of the world, UAV, and camera. The hunter drone will carry the stereo camera and net launcher system.} 
        \label{fig:frames}
    \end{center}
\end{figure}

To accomplish the detection of the intruder UAVs, a stereo camera provides useful information to identify the 3-dimensional position of objects. Based on the problem requirements, WithRobot oCamS-1CGN-U was selected for the study. This camera is supported by the Robot Operating System (ROS) framework. The camera has a depth range of 0.5 to 15 m, with a field of view of 92.8$\degree$. 

Throughout this paper, coordinate frames fixed to the world and the hunter UAV are referred, labeled as $W$ and $C$ respectively. Only for simplicity of notation, the hunter UAV and camera are fixed to each other and share the same coordinate frame.

\begin{figure}[b]
\begin{center}
\begin{subfigure}{.24\textwidth}
  \centering
  \includegraphics[width=0.95\linewidth, frame]{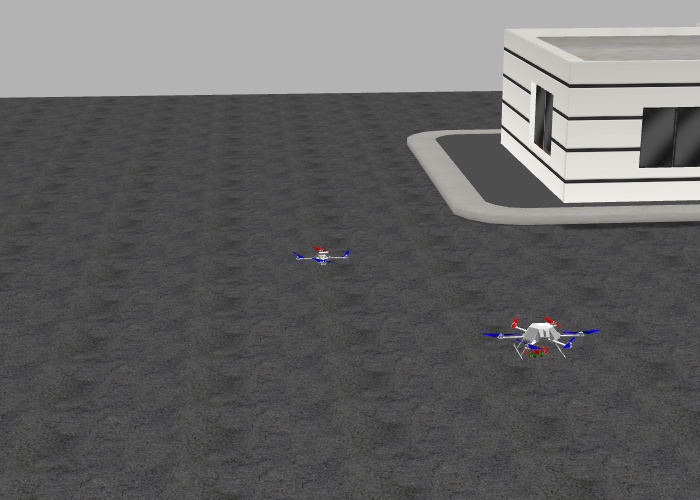}  
  \caption{}
  \label{fig:sub-first}
\end{subfigure}
\begin{subfigure}{.24\textwidth}
  \centering
  \includegraphics[width=0.95\linewidth, frame]{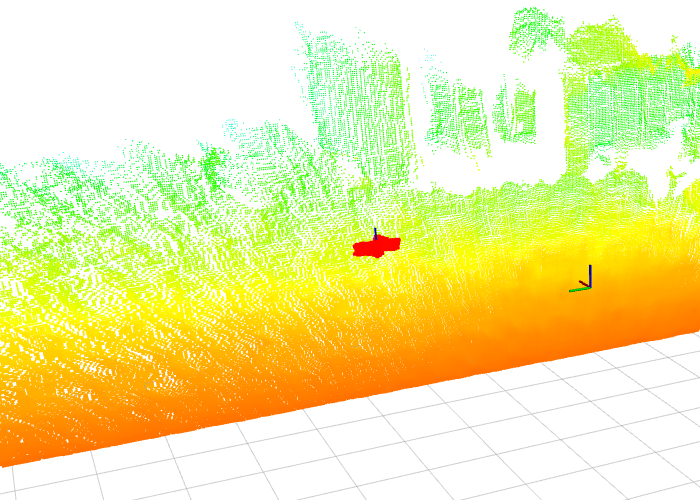}  
  \caption{}
  \label{fig:sub-second}
\end{subfigure}

\begin{subfigure}{.24\textwidth}
  \centering
  \includegraphics[width=0.95\linewidth, frame]{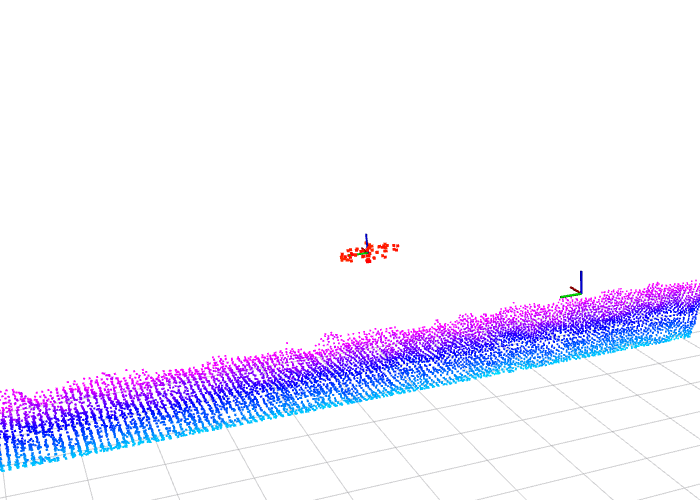}  
  \caption{}
  \label{fig:sub-third}
\end{subfigure}
\begin{subfigure}{.24\textwidth}
  \centering
  \includegraphics[width=0.95\linewidth, frame]{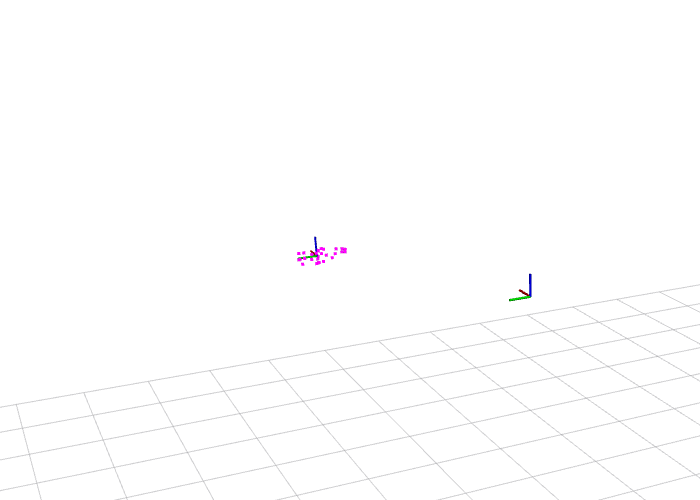}  
  \caption{}
  \label{fig:sub-fourth}
\end{subfigure}
\caption{Image processing pipeline. Grid 1x1 m. (a) Simulated environment, a hunter UAV and an intruder UAV. (b) Point cloud data obtained from the camera mounted on the hunter. (c) Filtered point cloud, down-sample  into  a  voxel  grid. (d) Cluster extraction. }
\label{fig:pipeline}
\end{center}
\end{figure}

The organization of the vision system for this application is briefly summarized in figure \ref{fig:pipeline}.

\subsection{3D Reconstruction from Stereo Vision}

The UAV targeting system uses a disparity algorithm to estimate the structure of the world in the form of a point cloud (i.e., a set of 3D points) \cite{hirschmuller_accurate_2005_2}. This algorithm uses a stereo camera, which is composed of two cameras displaced side to side. From the view-points of the two cameras, the disparity algorithm triangulates the 3D position of a point with the following equation:
\begin{equation}
    \begin{bmatrix}
        X \\
        Y \\
        Z
    \end{bmatrix} = \frac{b}{{}^{L}u-{}^{R}u}
    \begin{bmatrix}
        {}^{L}u \\
        {}^{L}v \\
        f
    \end{bmatrix}
\end{equation}

where $[X, Y, Z]^T$ is the point's 3D coordinate in the $C$ frame, and $u$ and $v$ are the horizontal and vertical pixel locations in the image. The focal length $f$, and baseline $b$ are intrinsic parameters of the camera.
The $R$ and $L$ left superscripts refer to the right and left camera, respectively.

The images were sampled at 30 FPS, with a resolution of 640 $\times$ 480 pixels, resulting in over 9 million data points every second (figure \ref{fig:sub-second}). To reduce the computational time of the algorithm, the point cloud was down-sampled into a voxel grid with a position density of 0.1 m \cite{rusu_semantic_2010}. In addition, the points beyond 10 m along the camera z-axis were discarded, since the 3D reconstruction for points past this distance is relatively inaccurate (figure \ref{fig:sub-third}).

\subsection{UAV Detection using Point Cloud Clustering}
A Euclidean clustering algorithm identifies clusters of points after the point cloud is filtered. Groups of points with at least a distance of 0.5 m from each other are considered different cluster groups. Clusters with more than 200 points or less than 20 points are removed since they likely do not represent the desired UAV (figure \ref{fig:sub-fourth}). 
The detections are selected as the centroid of each remaining cluster.

\subsubsection{UAV detection evaluation}
To evaluate the performance of the detection algorithm, the Precision-Recall curves were considered as in \cite{inproceedings}. The precision and recall metrics evaluate the performance in detecting all UAV appearances and in ignoring environmental or sensor noise, respectively.

\begin{equation}
    {Recall = \frac{TP}{TP+FN}, \hspace{8pt} Precision = \frac{TP}{TP+FP}}
\end{equation}
 where $TP$, $FN$, and $FP$ are the number of true positive, false negative, and false positive detections, respectively.

\subsection{Multi-Tracking of Possible UAVs}

A discrete Kalman filter estimates the state variables based on inaccurate and uncertain measurements. The Kalman filter consists of an iterative sequence of predictions of the current state ${x}_k$ from previous estimates $\hat{x}_{k-1}$, and corrections of the estimate with the detections $z_k$. The prediction stage considers the system state model (in this case constant acceleration), resulting in six dynamic parameters (position and velocity) and three measure parameters (position). 

Predict: $\hat{x}_{k}^{-} = A\hat{x}_{k-1} \hspace{10pt}$ Update: $\hat{x}_{k} = \hat{x}_{k}^{-} + K_k(z_k - H \hat{x}_{k}^{-})$

Where $A$ and $H$ are the Dynamics and Measuring matrices, $K$ is the Kalman gain (updated through $P$, $Q$ and $R$ which are the State error covariance, Process noise covariance, and Measurement noise covariance matrices).

The algorithm keeps multiple UAV tracks at any time and continuously adds or eliminates tracks depending on the incoming detected objects. New tracks are created when incoming detections are too far from all existing tracks. Similarly, deficient tracks are removed when they do not correspond to incoming detections after several consecutive iterations. 

This tracking framework has the advantages of 1) smoothing the trajectory of the track for stable aiming, 2) interpolating temporarily-undetected UAVs, 3) estimating the velocity and accelerations of the UAV, and finally, of 4) quantifying the statistical confidence of the UAVs' position, all using low computation resources \cite{sahbani_kalman_2016}.

The filter assumes a motion model of constant acceleration in 3D space, thus the track state is $X = [x, \dot{x}, \ddot{x}]^T$ (position, velocity and acceleration). In reality, the UAV track has a time-varying acceleration, so this variation was incorporated as an additive random noise to the state update equation. 

The Kalman filter uses noisy detection measurements to correct the state estimates. It assumes a 3D reconstruction error with a noise covariance matrix of $R^x = r$, where $r$ is a scaling factor. The state initialization error is modeled as a random variable with a covariance matrix of $P^x_0 = p \ diag\big(1^2, 30^2, 5.5^2\big)$, for a scaling value $p$. These numbers were selected since 1.0 $m$, 30 $m/s$, and 5.5 $m/s^2$ are reasonable maximum 3D reconstruction errors, UAV speed, and UAV acceleration. Note that $Q^x$, $R^x$, and $P^x_0$ are respectively the model noise covariance, measurement noise covariance, and the initial state error covariance matrices of a Kalman filter. The scaling parameters $q$, $r$, and $p$ are manually tuned based on the intruder UAV maneuverability and the stereo camera noise. 

\subsubsection{Data Association} 
Often there might be multiple active tracks at a moment in time and multiple incoming detections from a single camera frame. Using the linear sum assignment, detections are assigned to tracks by minimizing the sum of distances between the detections and their assigned tracks \cite{kuhn_hungarian_1955}. 
New tracks are created to accommodate detections that are farther to any other track by a maximum position error, $\epsilon_{max}$. 
 
If the position error exceeds $\epsilon_{max}$, the track is eliminated. 
Finally, the longest living track was selected for aiming after observing that the detections due to noise are transient and their corresponding tracks are quickly eliminated. 

\subsubsection{Performance evaluation metrics for video tracking}
For evaluation of the tracking algorithm, the metrics proposed by MOT16 were used as referred in \cite{milan_mot16_2016}. MOT16 is widely used for Multi-Object Tracking in videos, since it combines multiple sources of errors. Differently from the Recall and Precision metrics, it also accounts for the mismatch error Identity Switch (IDSW) (a target has been labeled with a different track number) with respect to the ground truth data $GT_k$. 
\begin{equation}
      {MOTA = 1 - \frac{\sum_k(FN_k + FP_k + IDSW_k)}{\sum_k(GT_k)}}
\end{equation}
 
\subsection{Position Control of the Hunter UAV}

From preliminary experiments it was found that the net launcher shoots the net with a parabolic trajectory. The net reaches a maximum expanded area when its center is located approximately 2 meters in front and 1 meter below the net launcher, or $x_E^d = [0, 1, 2]^T$ in the $C$ frame. Thus, the hunter UAV should control its own attitude to place the relative position of the moving intruder UAV at the desired $x^d_E$ location. 

\subsubsection{Position controller}
For the position control of the UAV, a dynamical model was developed \cite{lee_geometric_2010}. The goal is to achieve asymptotic tracking of four variables, three position variables for the center of mass of the UAV, and the direction of one body-fixed axis. This is applicable to this study by controlling the position of the hunter in x, z, and the rotation in the yaw axis, as to always keep the invader UAV in its field of view. The motion controller is implemented using the available implementation for the $lee\_position\_controller$ in the RotorS package for ROS.

\subsubsection{Desired Position} After the extraction of the best track, the hunter would know the position of the invader $x_E = [x, y, z]^T_C$ in its reference frame. The best location of the intruder to be captured is defined as $x^d_E$. The error in the position controller of the hunter is given by 
\begin{equation}
    {e_x = x_E^d - x_E}
    \label{error}
\end{equation}
The position control of the hunter was achieved sending a sequence of pose commands that set its trajectory to minimize the position error in equation \ref{error}.

\section{Evaluation of the UAV Targeting in a Computer Simulation}

\subsection{Error Analysis of the Detection}
A virtual environment for Micro Aerial Vehicle (MAV) simulation was set using the RotorS simulator \cite{inbook}. The stereo camera was modeled 
and an intruder and a hunter UAV are placed for testing.
 
The simulated environment provides ground truth information about the UAVs' position, which is used to compare with the results from the visual detection.

\subsubsection{Experiment} 
\begin{figure}[b]
    \begin{center}
        \includegraphics[width=5cm]{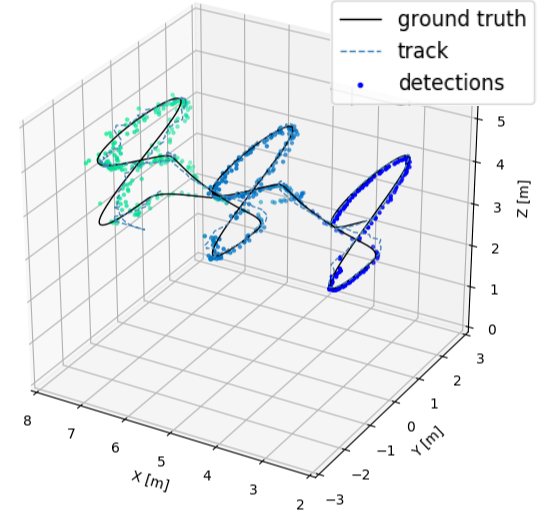}  
        \caption{Trajectory of the intruder UAV for the evaluation of detection and tracking accuracy.}
        \label{fig:3d}
    \end{center}
\end{figure}

An experiment was designed as follows: the stereo camera mounted on the hunter UAV sees an intruder UAV following a figure-8 trajectory at three different planes normal to the camera Z axis (figure~\ref{fig:3d}). The root mean square error (RMSE) and error analysis in detecting the intruder's centroid is shown in table~\ref{tb:mean}. This error analysis compared the ground truth against the detections and not to the tracking predictions.

\begin{table}[h]
\centering
  \captionsetup{width=\linewidth}
    \begin{center}
        \caption[width = \textwidth]{Spatial RMSE of detection in meters.}
        \begin{tabular}{cccccc}
            Distance  &      {} &    RMSE & {} &    Precision & Recall \\
            {[m]} &            X &     Y &     Z &     {} &        {} \\\hline
            2.5 &           0.063 &  0.112 &  0.032 &  1.000 &      0.974 \\\hline
            5.0 &           0.073 &  0.108 &  0.035 &  1.000 &      0.983 \\ \hline
            7.5 &           0.174 &  0.083 &  0.032 &  0.998 &      0.993 \\\hline
        \end{tabular}
        \label{tb:mean}
    \end{center}
\end{table}

\subsection{Execution time of detection} 
A detection lag of $\Delta t_{Det} = 0.15s$ was calculated as the maximum cross-correlation coefficient between the ground truth and detection curves (figure~\ref{fig:zoom}). . Similarly, the tracking lag with respect to the ground truth was $\Delta t_{Track} = 0.15s$. This means that the computational time of the detection is the limiting part of the algorithm.

\subsection{Tracking Accuracy of the Intruder UAV}
\begin{figure}[t]
    \begin{center}
        \includegraphics[width=8cm]{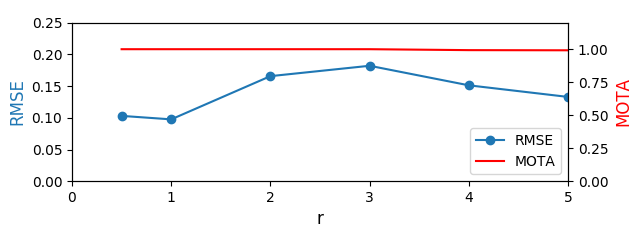}  
        \caption{Root mean square error of the best track when compared to the ground truth data, for different values of $r$ in the Kalman filter.}
        \label{fig:mota}
    \end{center}
\end{figure}

In figure~\ref{fig:mota} the MOTA and RMSE was evaluated for several  $r$ parameters, ranging from 0.5 to 5.0. The $r$ parameter scales the measurement noise covariance matrix of the Kalman filter supporting the tracking algorithm. For all tested values of $r$, the MOTA was above 0.999 and the RMSE had a minimum value of $0.09$ meters at $r = 1.0$. Note that for very small values of $r$ the tracks rely more on the noisy measurement (detection) data, producing a noisy track. While for very large values of $r$, the tracking algorithm trusts the motion model too much, and does not update the acceleration state in a timely manner. 

\begin{figure}[b]
    \begin{center}
        \includegraphics[width=6.5cm]{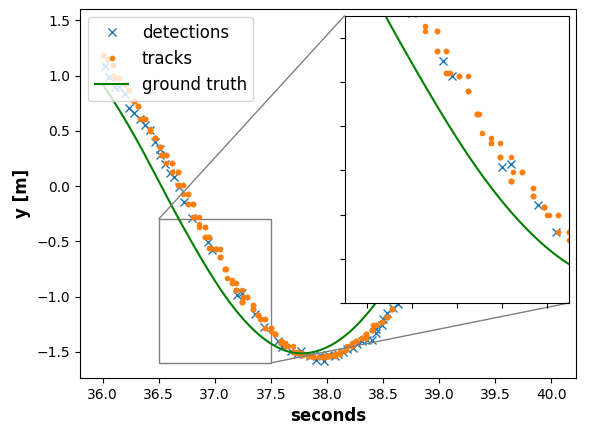}  
        \caption{Part of the trajectory of the intruder in the y axis.}
        \label{fig:zoom}
    \end{center}
\end{figure}

An advantage of the Kalman filter approach is shown in figure \ref{fig:zoom}, where the tracking algorithm correctly predicts the position of the intruder UAV even when there are temporarily missing detections.

\section{Discussion}
Using point cloud data from a stereo camera as the rubric for UAV recognition shows to be a feasible solution. A good recall index is needed to guarantee the detection of all invader UAVs. Table ~\ref{tb:mean} shows a reliable recall index, and a precision that stays high as recall increases. 

The tracking algorithm based on a Kalman filter has the smoothing and predictive characteristics required for a steady aiming, as shown in figure~\ref{fig:zoom}. The detection and tracking lags were similar, suggesting that the detection step has a longer execution time. 

A value of MOTA close to 1.0 means the tracking algorithm always assigned the intruder UAV to the same track, even during missing detections. This is important for target aiming applications to avoid switching between targets. Additionally, the tracking error of 0.09 meters $RMSE$ is considerably smaller than the area of the net, which spans above 1 meter in diameter. Therefore, this tracking error is small enough for a successful intruder UAV capture.

Future work will focus on adapting and testing the algorithm on a real-world scenario, designing an optimal motion controller to aim at the invader UAV that will compensate for the tracking lag and the ballistic dynamics of the net launching.

\section{Conclusion}
This paper proposes a drone hunting platform capable of detecting invader UAVs in an area using stereo vision. The algorithm includes an object detection step based on computer vision and a tracking step based on a Kalman filter. This algorithm is can be executed in real-time on an onboard computer and mounted on a commercial UAV platform. The fast execution time of the algorithm added to the predictive nature of the tracking algorithm, make this a reliable system for the detection of intruder UAVs for use in aerial security.

\bibliography{ifacconf}

\end{document}